\documentclass[prd,aps]{revtex4}

\begin{document}
\title{Uniform discretizations: a quantization procedure for
totally constrained systems including gravity}

\author{ Miguel Campiglia$^{1}$, Cayetano Di Bartolo$^{2}$ 
Rodolfo Gambini$^{1}$,
Jorge Pullin$^{3}$}
\affiliation {
1. Instituto de F\'{\i}sica, Facultad de Ciencias,
Igu\'a 4225, esq. Mataojo, Montevideo, Uruguay. \\
2. Departamento de F\'{\i}sica,
Universidad Sim\'on Bol\'{\i}var, Aptdo. 89000, Caracas 1080-A,
Venezuela.\\
3. Department of Physics and Astronomy, Louisiana State University,
Baton Rouge, LA 70803-4001}

\begin{abstract}
We present a new method for the quantization of totally constrained
systems including general relativity. The method consists in
constructing discretized theories that have a well defined and
controlled continuum limit. The discrete theories are constraint-free
and can be readily quantized. This provides a framework where one can
introduce a relational notion of time and that nevertheless
approximates in a well defined fashion the theory of interest. The
method is equivalent to the group averaging procedure for many systems
where the latter makes sense and provides a generalization
otherwise. In the continuum limit it can be shown to contain, under
certain assumptions, the ``master constraint'' of the ``Phoenix
project''. It also provides a correspondence principle with the
classical theory that does not require  to consider the semiclassical
limit.
\end{abstract}

\maketitle
The issue of the dynamics is perhaps the central problem in
canonical quantization approaches to totally constrained theories like
quantum general relativity \cite{problemofdynamics,phoenix}. There are three
salient aspects of the problem that have prevented from advancing in
the quantization. The first one is how to construct a space of
physical states for the theory that are annihilated by the quantum
constraints and that is endowed with a proper Hilbert space
structure. The second issue is related to the introduction of a
correspondence principle with the classical theory, in particular to
check the constraint algebra at a quantum level. The third problem is
how to address the ``problem of time'' \cite{kuchar} that is, to introduce a
satisfactory picture for the dynamics of the theory in terms of
observable quantities.

We have proposed in previous papers \cite{paradigm} a paradigm to deal
with the above issues that consists in describing the theory in terms
of a discrete evolution. This is analogous for instance to lattice
QCD, where one uses a discrete theory to approximate and define the
continuum quantum theory as a suitable limit.  In our approach the
discretization is carried out in such a way that the dynamics of the
discrete theory is unconstrained. We call this approach ``consistent
discretizations¨''. The lack of constraints in the discrete theory
bypasses almost automatically the three issues mentioned above for the
discrete theory, but leaves open the problem of how to define a
continuum limit. One of the important points of our proposal is that
the discretization is carried out in a way that the discretization
step is dynamically determined and therefore one does not have a
direct control on how to take the continuum limit as one does in
ordinary discretizations.

In this article we want to spell out a constructive technique to
define properly the quantum continuum limit. Our procedure can
therefore be viewed as an alternative to the Dirac quantization
procedure for a continuum theory in the sense that at the end of the
day it yields a quantum continuum theory. We will see that the
procedure has attractive advantages with respect to the Dirac
procedure.

A key element is the introduction of a set of discretizations for a
given continuum theory called ``uniform discretizations''. These are
such that the evolution steps are bounded by a value that one chooses
in the initial data. The value of the constraints of the continuum
theory (which are not exactly satisfied in the discrete theory) are
also bounded throughout the evolution. This is of interest in itself
since it is quite non-trivial to find discretizations of general
relativity for which the constraints remain bounded. The important
aspect is that since one controls the discretization step through the
initial data, one can define properly a continuum limit just by
choosing data that has a step as small as is desired.

We start by considering a canonical theory with a phase space with
canonical variables $q_i,p_i$ that is totally constrained, by this
meaning that the total Hamiltonian $H_T$ is a linear combination of $N$
constraints $\phi_i$ which we will assume are first class. We are
assuming we are dealing with a mechanical system with a finite number
of degrees of freedom. This is of interest in the context we are
discussing since field theories when formulated on the discrete space
---as is common, for instance, in loop quantum gravity---, become such
systems.

We will now introduce a discrete evolution given by the flux of a
Hamiltonian $H$ that is constructed from the constraints of the theory
in a way we will soon discuss and such that the evolution of any
dynamical variable $A$ is given by
\begin{equation}
A_{n+1} = e^{\{\bullet,H\}}(A_n)\equiv A_n +\{A_n,H\}
+{1 \over 2}\{\{A_n,H\},H\}+\cdots
\end{equation}
As is obvious, $H$ is a constant of the motion of the discrete evolution.

The uniform discretizations are given by a family of Hamiltonians $H$
constructed in the following way. Consider a smooth function of $N$
variables $f(x_1,\ldots,x_n)$ such that the following three conditions
are satisfied: a) $f(x_1,\ldots,x_n)=0 \iff x_i=0 \forall i$ and
otherwise $f>0$; b) ${\partial{f}\over\partial{x_i}}(0,\ldots,0)=0$;
c) ${\rm det} {\partial^2 f \over \partial x_i \partial x_j}\neq 0$
$\forall x$ and d) $f(\phi_1(q,p),\ldots
\phi_N(q,p))$ is defined for all $q,p$ for the complete phase space. 
Given this we define $H(q,p)\equiv f(\phi_1(q,p),\ldots
\phi_N(q,p))$.

A particularly simple example is $H(q,p) ={1/2}\sum_{i=1}^N \phi_i(q,p)^2$,
a choice that has interesting parallels with the ``master constraint''
of the ``Phoenix project'' \cite{phoenix} as we shall discuss later.

An important point is that if we choose initial data such that
$H<\epsilon$ then $\phi_i$ remain bounded throughout the evolution and
will tend to zero in the limit $\epsilon\to 0$. Let us see that in
this limit one recovers the evolution equations given by the total
Hamiltonian $H_T$ in the constrained continuum theory. Let $H$ as in
the simple example above and take its initial value to be
$H_0=\delta^2/2$. We define $\lambda_i=\phi_i/\delta$, and therefore 
$\sum_{i=1}^N \lambda_i^2 =1$. The evolution of the dynamical
variable $q$ is given by,
\begin{equation}
q_{n+1}=q_n +\sum_{i=1}^N\{q_n,\phi_i\} \lambda_i\delta +O(\delta^2)
\end{equation} 
and if we define $\dot{q}\equiv \lim_{\delta\to
0}(q_{n+1}-q_n)/\delta$, where we have identified the ``time
evolution'' step with the initial data choice for $\delta$, one then
has,
\begin{equation}
\dot{q} =\sum_{i=1}^N\{q,\phi_i\}\lambda_i, 
\end{equation}
and similarly for other dynamical variables. The specific values of
the multipliers $\lambda_i$ depend on the initial values of the
constraints $\phi_i$.

Notice that in the current proposal the evolution is generated by the
exponential of a Hamiltonian. In our previous treatments of
``consistent discretizations'' one used as a starting point a type 1
or 2 generating function of a canonical transformation. Generically
one can associate a generating function with a choice of $f$, by
constructing the Hamilton principal function. The knowledge of such
generating function could be useful, for instance, to actually use the
discrete scheme in a concrete numerical implementation classically. As
an example, we shall analyze a system with $N$ Abelian constraints by
using a slightly different technique that seems more convenient for
numerical applications. We define
\begin{equation}
L(q_n,q_{n+1},\lambda_1,\ldots,\lambda_N)=
S(q_n,q_{n+1}\lambda_1,\ldots,\lambda_N)
+g(\lambda_1,\ldots,\lambda_N)
\end{equation}
where $L$ is a type 1 generating function of a canonical
transformation between canonical variables $q_n,p_n$ and
$q_{n+1},p_{n+1}$, $S$ is Hamilton's principal function for a given
set of Lagrange multipliers $\lambda_1,\ldots,\lambda_N$ (they are
evaluated at instant $n$, we omit the subscript for simplicity) and
$g$ is a smooth function such that $g(0)=0,\partial_i{g(0)=0}$ and
${\rm det} {\partial^2 g \over \partial x_i \partial x_j}\neq 0$. The
generating function yields the canonical momenta in the usual way
$p_{n+1}=\partial L /\partial q_{n+1}$, $p_n=-\partial L /\partial
q_n$. One also has that $\partial L/\partial \lambda_i=0$. The latter
equation determines the Lagrange multipliers,
$\lambda_i=h_i(\phi)$,
where $h_i$ is the inverse function of the mapping defined by
$\lambda_i\rightarrow {{\partial{g}}\over{\partial{\lambda_i}}}$.
This evolution corresponds to a Hamiltonian
$H=f(\phi_1,\ldots,\phi_N)$, with $\partial_i{f}=h_i$. In other words,
the Legendre mappings induced by $g$ and $f$ are inverse. In
particular, when $g=\sum_{i=1}^N x_i^2/2$ then $H=\sum_{i=1}^N
\phi_i^2/2$. 
The generating function $L$ allows to determine the discrete evolution
that preserves exactly the value of the constraints of the continuum
theory and recovers the continuum limit when all $\phi_i\to 0$ in the
initial data.

The constants of the motion of the discrete theory are quantities that
have vanishing Poisson bracket with the Hamiltonian, $\{O^D_i,H\}=0$
and in the continuum limit $H_0 \to 0$ reproduce, as functions of
phase space the ``perennials'' of the continuum theory
$O^C_i=\lim_{H_0\to 0} O^D_i$. This can be immediately seen from the
fact that the discrete equations reproduce the continuum equations for
any dynamical variable in the continuum limit. Conversely, for every
perennial of the continuum theory there exists a constant of the
motion (in general many constants) of the discrete theory that reduce
to the given perennial in the continuum limit.
We have therefore shown that uniform discretizations recover the constraints
and the perennials of the continuum theory and therefore provide a good
starting point for a quantization of the continuum theory.

We now turn our attention to the quantum theory. We will introduce a 
Heisenberg quantization for the discrete theory (this is more natural
given that one has an explicit evolution).
To quantize the theory we follow several steps. We start with the classical
discrete system constructed as in the previous section, we eliminate the
canonical variables at level $n+1$ in terms of the variables at level $n$,
$
q_{n+1} = q_{n+1}(q_n,p_n), \quad p_{n+1} =p_{n+1}(q_n,p_n)$.

We then define the
kinematical space of states of the quantum theory, ${\cal H}_k$, as 
the space of functions of $N$ real variables $\psi(q)$ that are square
integrable. In this space we define operators $\hat{Q}$ and $\hat{P}$
as usual.
To construct the operators at other time levels (in the Heisenberg
Picture) we introduce a linear invertible operator $\hat{U}$ that we will
define later and we take
\begin{equation}
\hat{Q}_{n} \equiv \hat{U}^{-1} \hat{Q}_{n-1} \hat{U} = 
\hat{U}^{-n} \hat{Q}_0
\, \hat{U}^n\,, \quad \hat{P}_{n} \equiv \hat{U}^{-1} 
\hat{P}_{n-1} \hat{U} =
\hat{U}^{-n} \hat{P}_0 \,\hat{U}^n\,.
\end{equation}

When the evolution is determined by a discrete Hamiltonian $H$, as
is the case in the uniform discretizations, the evolution operator is
given by $\hat{U}=e^{-i\hat{H}/\hbar}$. Notice that $\hat{U}$ may also
be determined by requiring that the fundamental operators satisfy an
operatorial version of the evolution equations,
\begin{equation}\label{Cuant-005}
\hat{Q}_{n}\, \hat{U} - \hat{U} Q_{n+1}(\hat{Q}_n,\hat{P}_n)=0, 
\quad \hat{P}_n\, 
\hat{U} - \hat{U}
P_{n+1}(\hat{Q}_n,\hat{P}_n)=0,
\end{equation}
and this provides a consistency criterion for the construction of $\hat{U}$.

At a classical level $H=0$ if and only if the constraints $\phi_i=0$.
There exists a natural definition of the physical space of the
continuum theory that does not require that we refer to the
constraint. Since we know that $\hat{U}=\exp(-i \hat{H}/\hbar)$, a necessary
condition satisfied by the states of the physical space of the
continuum theory, $\psi\in{\cal H}^c_{\rm phys}$ is given by $\hat{U}\psi =
\psi$. More precisely the states $\psi$ of  ${\cal H}^c_{\rm phys}$ should 
belong to the dual of a space $\Phi$ of functions sufficiently 
regular on ${\cal H}_{\rm phys}$.
That is, the states $\psi\in {\cal H}^c_{\rm phys}$ satisfy 
\begin{equation}
\int \psi^* \hat{U}^\dagger \varphi dq = \int \psi^* \varphi dq,
\end{equation}
where $\varphi\in \Phi$. 
This condition characterizes the quantum physical space 
of a constrained continuum theory without needing to implement the 
constraints as quantum operators by using the discretization technique.

The unitary operators of the discrete theory allow to
construct the ``projectors'' onto the physical space of the continuum
theory, which is one of the main goals of any quantization procedure
based on Dirac's ideas. It should be noted that these are really 
generalized projectors in the sense that they project to a set of functions
that belong in the dual of a subspace of sufficiently well behaved
functions of ${\cal H}_k$. All of this is achieved without having to
define the quantum constraint. To construct the ``projectors'' one can
compute,
\begin{equation}
\hat{P}\equiv \lim_{M\to\infty} C_M \hat{U}^M.
\end{equation}
If such a limit exists for some $C_M$ such that 
$\lim_{M\to\infty}(C_{M+1}/C_M)=1$ then $\hat{U} \hat{P} = \hat{P}$, and 
we have that
$\hat{U} \hat{P} \psi = \hat{P}
\psi$,  $\forall \psi \in {\cal H}_{\rm k}$.

The limit exists in several examples in which H has a continuum
spectrum, as we shall see. If the spectrum is discrete with
eigenvalues $e_i$ and it contains a vanishing eigenvalue $e_{i_0}$
then a projector is trivially defined as $|e_{i_0}>< e_{i_0}|$. A
constructive procedure leading to a general definition of the
projector in terms of the discrete evolution operator $\hat{U}$ valid for
any spectrum, continuum or discrete, is given by:
\begin{equation}
\hat{P}\equiv \lim_{M\to\infty} \sum_{n=M}^{{\rm Int}(rM)}
{{C_n \hat{U}^n}\over{{\rm Int}(rM)-M}}.
\end{equation}
where $r$ is a real number grater than one and ${\rm Int}(rM)$ is the
integer part of $rM$. If U has a continuum spectrum this definition is
a trivial consequence of the previous one. In the case of a discrete
spectrum one can check that the definition works recalling the
definition of the Kronecker delta in terms of a Fourier series.
Notice that the definition of physical space that Thiemann introduces
in the ``phoenix project'' \cite{phoenix}, is equivalent to the choice
we make if one is considering the Hamiltonian that is quadratic in the
constraints.
Furthermore, given two states of ${\cal H}_{\rm phys}$, $\psi_{\rm ph}$,
$\phi_{\rm ph}$, where $\psi_{\rm ph}=\hat{P} \psi$, and
$\phi_{\rm ph}=\hat{P} \phi$, the physical inner product
is defined by $<\psi_{\rm ph}|\phi_{\rm ph}>=
\int dq \phi(q)^* \hat{P} \psi(q)$ and a physical inner
product is  determined by  the projector constructed from 
the discrete theory.

We now illustrate the technique with a rather general example. We consider
a generic mechanical system with a finite dimensional phase space
with one constraint $\phi=0$. We will show that the projector 
constructed with our technique reproduces the one constructed with
group averaging techniques \cite{groupaveraging}. That is,
\begin{equation}
P =\lim_{M\to \infty} \sqrt{4 \pi i M} e^{i M \phi^2} = \int_{-\infty}^\infty
{d\mu \over 2 \pi} e^{i \mu \phi}.  
\end{equation}
To make contact with the group averaging case we need to 
assume that $\phi$ is a self-adjoint operator in the kinematical
phase space with an eigenbasis given by 
$\phi |\alpha> = \phi(\alpha) |\alpha>$ and $1=\int |\alpha><\alpha| d\alpha$.
The proof of the equivalence is,
\begin{equation}
P =P \int |\alpha><\alpha| d\alpha = \int \lim_{M\to\infty} \sqrt{4 \pi i M}
e^{-i M \phi^2} |\alpha><\alpha|
\end{equation}
and noting that $\lim_{M\to \infty} \sqrt{4 i \pi M} e^{-i M x^2} =\delta(x) = 
\int_{-\infty}^{\infty} {d \mu \over 2 \pi}e^{i\mu x}$ the proof is complete.
For the proof we assumed a quadratic form of the Hamiltonian, but can
actually be extended to Hamiltonians of the general form we discussed above,
computing the integral by steepest descents. The proof can also be extended
to systems with $N$ Abelian constraints by noting that $\lim_{M\to \infty} 
({4 i \pi M})^{N/2} e^{-i M {\vec x}.{\vec x}} =\delta(\vec x) = 
\int_{-\infty}^{\infty} {d \mu^N }e^{i{\vec \mu}.{\vec x}}/(2 \pi^N)$. 
This includes important cases
like gravity in $2+1$ dimensions, where one can immediately reproduce
the results obtained by Perez and Noui \cite{pereznoui} via group
averaging.

Summarizing, the method of uniform discretizations allows to tackle
satisfactorily the three central problems of the dynamics of quantum
general relativity and provides new avenues for studying numerically
classical relativity as well. It is based on a set of discretizations
generated by Hamiltonians that contain as a particular case the
quadratic Hamiltonian of Thiemann's ``master constraint
programme''. The phase space of the continuum theory and the physical
inner product can be constructed in a straightforward way from the
discrete theory and therefore provides a generalization of the group
averaging extension of the Dirac procedure to systems with structure
functions in their constraint algebra, like is the case in general
relativity. The use of non-quadratic Hamiltonians is possible and adds
flexibility to the method. The flexibility is crucial, for instance in
tackling in an extremely compact and straightforward way gravity in
$2+1$ dimensions (we will discuss this example and other issues
in a forthcoming paper). In other approaches to the dynamics of quantum
gravity a major obstacle is the need to define the constraints as
quantum operators in an unambiguous way.  This may be due to the fact
that the constraints are only well defined on a diffeomorphism
invariant space of sets where the constraint algebra is trivial, or,
as in the case of the ``master constraint'' since one has only one
constraint. This requires an a-posteriori study of each quantization
proposal to determine if they can reproduce general relativity in some
suitable semi-classical limit. This is complex and difficult to carry
out. One is therefore left with proposals that one is not even sure if
they have any connection with the theory one desires to quantize. In
the discrete approach $\hat{U}=e^{-i \hat{H}/\hbar}$ must implement the
discrete classical evolution associated to the canonical
transformations and therefore one has a correspondence constructive
principle as a guide that requires that the quantum evolution
equations reproduce the Heisenberg equations associated with the
classical theory. Contrary to other methods, one can construct the
continuum theory as approximated by a discrete theory in the kinematical
Hilbert space. This allows the use of operators that can be genuinely
used as quantum mechanical clocks and is therefore possible to characterize
the evolution in a relational way in terms of conditional probabilities
\cite{njp}. The method is therefore devoid of the usual conceptual problems of 
canonical quantum gravity.

In conclusion, just like lattice methods did for QCD, the uniform
discretizations transform the quantization of constrained systems into a
computational exercise.  The only major hurdle that could stand in the
way is that the continuum limit may not exist for the case of full
general relativity. In that case, this will be a strong indication
that the theory does not exist.  The method allows to incorporate all
the benefits of the kinematics of loop quantum gravity \cite{loop} and
provides an unambiguous avenue to characterize the dynamics and
complete the quantum theory.

This work was supported in part by grants NSF-PHY-0244335,
NSF-PHY-0554793, CCT-LSU, Pedeciba, DID-USB grant GID-30 and Fonacit
grant G-2001000712 and the Horace Hearne Jr. Institute for Theoretical
Physics.

\end{document}